# In situ synthesis of polyynes in a polymer matrix by pulsed laser ablation in liquid

Sonia Peggiani[a], Anna Facibeni[a], Alberto Milani[a], Chiara Castiglioni[b], Valeria Russo[a], Andrea Li Bassi[a], Carlo S. Casari[a],*

[a]Department of Energy, Politecnico di Milano, Via Ponzio 34/3, 20133 Milan, Italy
[b]Department of Chemistry, Materials, and Chemical Engineering "Giulio Natta", Politecnico di Milano, Piazza Leonardo da Vinci 32, 20133 Milano, Italy

## *Abstract*

Polyynes are finite chains formed by sp-hybridized carbon atoms with alternating single and triple bonds and displaying intriguing electronic and optical properties. Pulsed laser ablation in liquid (PLAL) is a well assessed technique for the physical synthesis of hydrogen-capped polyynes in solution, however, their limited stability prevents further exploitation in materials for different applications. In this work, polyynes in poly(vinyl alcohol) (PVA) were produced in a single-step PLAL process by ablating graphite directly in aqueous solution of PVA, investigating the role of polymer concentration. The presence of PVA solution, as a participating medium for PLAL, is shown to favour the formation of polyynes. The addition of Ag colloids to the aqueous PVA/polyynes solution allowed surface-enhanced Raman spectroscopy (SERS) measurements, carried out both on liquid samples and on free-standing nanocomposites, obtained after solvent evaporation. We show that polyynes in the nanocomposite remain stable at least for 11 months, whereas the corresponding PVA/Ag/polyynes solution displayed a strong polyyne reduction already after 3 weeks. These results open the view to further characterizations of the properties of polyyne-based films and materials.

*Corresponding author. Email: carlo.casari@polimi.it (Carlo Casari)
Phone number: +390223996331

# Introduction

Last achievements in the research on carbyne and linear sp-carbon chains have outlined great potential for their huge effective area, the largest Young modulus, high electron mobility and thermal conductivity [1-5]. A few proof of concept demonstrations of possible applications have been reported only recently, as in the case of optical imaging of biomolecules and field effect transistor realization [6, 7]. Finite sp-carbon linear chains with single-triple bond alternation, also called polyynes, display semiconducting optoelectronic properties modulated by the length (i.e. number of carbon atoms) and by the terminations (atom or molecular group) [8]. Polyynes can be synthesized by physical methods potentially scalable to mass production such as, for example, submerged arc discharge in liquid (SADL) and pulsed laser ablation in liquid (PLAL). In both cases, the most exploited liquid media are organic solvents [9-14], whereas the use of water is less explored in literature [15-17]. However, it was recently showed the use of water in SADL to reach polyynes up to 16 carbon atoms or to collect the gas generated by the process [18, 19].

To fully take advantage of the properties of polyynes, the availability of a stable material in ambient conditions is required. Stability of sp-carbon atomic chains is not a trivial issue [20]. In fact, hydrogen-capped polyynes, typically synthesized by physical methods such as SADL and PLAL, show poor stability against cross linking, especially when the solvent is removed [21-23]. Different strategies have been developed so far to avoid degradation, as for instance encapsulation of long sp-carbon wires inside double- or multi-walled carbon nanotubes [24, 25], or when polyynes are surrounded by a rotaxane macrocycle [26] or immersed in an ionic liquid [27]. In general, bulky groups at the end of the sp-carbon chains are known to act as spacers preventing cross-linking reactions [28] as, for example, $sp^2$ clusters [29], graphene edges [30] or phenyl groups [8, 31-33]. Another very promising technique to stabilize carbon chains is to embed them in solid matrices, such as solid Ag nanoparticle assemblies [22], $SiO_2$ dried gel [34] and poly(vinyl alcohol) (PVA). PVA is widely exploited in the formation of carbon or metal-based composites because it is a low-cost material, chemically stable, soluble in water with good film formation properties [35, 36]. A few works have addressed the encapsulation of polyynes in such a polymer to form a nanocomposite by simply adding PVA to a polyynes solution [37-39]. In view of possible future applications, further investigations are needed to understand structure-property relationship and to improve easy and fast production methods.

In this paper, a one-step and simple method for in situ synthesis of polyynes in PVA is presented by employing laser ablation of graphite in aqueous solutions of PVA. In this way, it is possible to achieve a two-fold advantage, on one side the use of non-toxic liquids as solvents for polyynes synthesis and on the other, no need of any further heating or any dilution to obtain a solution of PVA and polyynes. The participating role of aqueous solutions of PVA at different concentrations as a solvent for PLAL is explored by UV-Vis and surface-enhanced Raman spectroscopy (SERS). After adding Ag colloid to the aqueous solutions of PVA with polyynes, the solution was left to dry forming a free-standing film. SERS analysis in different areas of the film was performed to investigate the uniformity of polyynes dispersion and to evaluate the connection between morphology and the SERS enhancement. Finally, a comparison between the stability of PVA/Ag/polyynes solution and the corresponding free-standing film is given, confirming the improvement of the stability of polyynes embedded in the polymer.

## 1. Experimental procedures

Polyynes were synthesized by ablating a graphite target in liquid media by means of ns-pulsed laser. Ablation was performed by a Nd:YAG laser (2$^{nd}$ harmonic, $\lambda$ = 532 nm) with pulse duration of 5–7 ns and repetition rate of 10 Hz (Quantel Q-smart 850). The laser beam was focused from the top side of the vial. The estimated energy fluence reaching the target was 5.2 J/cm$^2$. The solvent was prepared dissolving poly(vinyl alcohol)(PVA) granules (Fluka, MW 130 000, degree of polymerization $N \sim$ 2700) in deionized water Milli-Q (0.055 μS) at 373 K and at different concentrations, namely 0.03, 0.5, 1, 3, 10 wt.%. We hereafter name the samples as PVAx%. A graphite target 2 mm thick and 8 mm in diameter with a purity of 99.99% (Testbourne Ltd) was placed at the bottom of the glass vial containing 10 ml of solution. To perform surface-enhanced Raman spectroscopy (SERS), we varied the ratio of polyynes in aqueous PVA solution (indicated by PVA/polyynes) with respect to silver nanoparticles to obtain the highest SERS intensity of polyynes. We achieved the optimum for a volume ratio of 1:2 (1 for PVA/polyynes, 2 for Ag colloid). We refer to the SERS-active solution as PVA/Ag/polyynes solution. The Ag nanoparticles, employed in this work, were made by Lee-Meisel method [40], reaching a concentration of 10$^{-3}$ M and a plasmonic peak centred at 413 nm. Their medium size of 60–80 nm was estimated thanks to the morphological characterization by the field emission scanning electron microscope (FEG-SEM, Zeiss Supra 40). The drop-casted solution of PVA/Ag/polyynes was left to dry in a small plastic container at room temperature. After 24 hours, we detached the solid PVA/Ag/polyynes nanocomposite material, and we squared it adding a graduate scale to perform localized SERS analyses. The step by step procedure for the formation of free-standing films is reported in Table 1.

**Table 1**

Steps for the preparation of polyynes-based nanocomposite

|  | Description | Sample Name |
|---|---|---|
| 1st step | Preparation of solvents for PLAL, i.e. PVA in different concentration in water wt.%: x = 0.03, 0.5, 1, 3, 10 | PVAx% |
| 2nd step | PLAL of graphite in aqueous PVA at different concentration. | L-x% |
| 3rd step | Addition of Ag colloid in a volume ratio of 2:1 with respect to PVA/polyynes solutions. | L-x%_Ag |
| 4th step | Drop casting and solidification for 24 hours at room temperature. | S-x%_Ag |

UV-Vis absorption spectra of PVA/polyynes liquid samples were obtained by Shimadzu UV-1800 UV/Visible Scanning Spectrophotometer (190-1100 nm). SERS measurements were carried out using a Renishaw InVia micro Raman spectrometer with an argon ion laser ($\lambda$=514.5 nm) and a spectral resolution of about 3 cm$^{-1}$. SERS spectra of both PVA/Ag/polyynes solutions and the corresponding nanocomposites were taken with an objective of 50X and a laser power at the sample of 0.13 and 0.013 mW, respectively on liquid and solid one. SERS maps of PVA1%/Ag/polyynes film were carried out by a Renishaw InVia micro Raman spectrometer equipped with an automatic translator x-y-z of micrometric resolution and a diode-pumped solid-state laser ($\lambda$=532 nm), using a laser power of 0.35 mW. The laser powers set in these experiments do not cause any laser-induced degradation to the material.

## 2. Results

Laser ablation of a solid graphite target was directly performed in aqueous solutions of PVA to achieve in situ synthesis of hydrogen-capped polyynes in a polymer. UV-Vis absorption spectra of solutions with different concentration of PVA in water (e.g. 0.03, 0.5, 1, 3, 10 wt.%) before and after ablation of graphite are shown in **Figure 1**. We referred to the PVA/polyynes solutions by L-0.03%, L-0.5%, L-1%, L-3%, L-10%, whereas to the ablation in water by L-0%. Pure PVA is transparent in the visible range and it has an absorption peak at 278 nm, related to the $\pi \rightarrow \pi^*$ transition of the carbonyl groups (C=O), associated to the ethylene unsaturation (C=C) of the type $-(CH=CH)_2CO-$, typical of end groups, as reported in Ref. [41]. By increasing the concentration of PVA in water, a consequent growth of the absorption peak at 278 nm is observed. After ablation, UV-Vis spectra indicate the presence of polyynes by the characteristic vibronic peaks at 199, 206, 215, 225 and 251 nm. These features are related to H-terminated polyynes, which for simplicity we indicate as $C_n$, with n (i.e. the number of carbon atoms) ranging from 6 to 10 in agreement with previous works [18, 42, 43]. Longer polyynes may be present but hardly detectable, due to their low concentration, which results in a weak signal also partially covered by the PVA peak. In fact, for PVA concentration exceeding 0.5 wt.%, the strong PVA peak makes it difficult to identify the presence of polyyne. By comparing the spectrum of L-0.03% with L-0% after proper subtraction of the corresponding PVA background (see inset of **Figure 1**), we observe a higher absorption of polyynes synthesized in PVA0.03%. By the peak-valley difference in absorbance of the polyynes 0-0 band of the $^1\Sigma_u^+ \leftarrow X^1\Sigma_g^+$ transition, we register an increase of a factor 2.3 both for $C_6$ (at 199 nm) and $C_8$ (at 225 nm). This indicates that aqueous PVA, employed as a PLAL solvent, favours the formation of polyynes with respect to pure water. This effect can be associated to the presence of PVA and may help the sp-carbon chains formation by providing further carbon and hydrogen atoms [43] and by changing the viscosity of the solution, as discussed below.

Then, we added chemically synthesized Ag colloid to the PVA/polyynes solutions to characterize the sample composition and structure by surface enhanced Raman spectroscopy (SERS). Finally, we prepared the solid nanocomposite after drop casting the liquid specimen on a substrate and leaving it to dry at room temperature in stationary conditions.

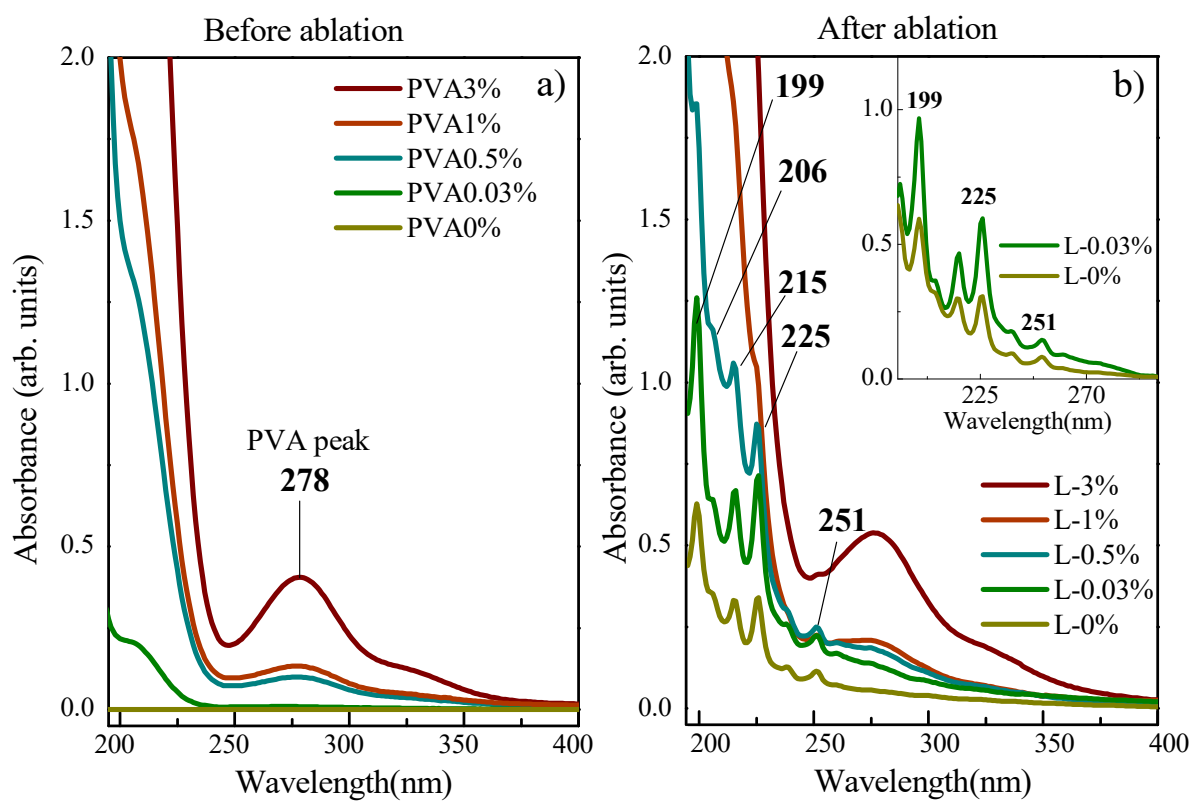

*Figure 1. UV-Vis spectra of PVA solutions at different concentrations in water **a)** before and **b)** after ablation. Inset: comparison of spectra of polyynes solution with and without PVA0.03%. Spectra are obtained after subtraction with the corresponding background spectra.*

Raman spectroscopy can easily distinguish the signal of polyynes from the one of PVA. In fact, the sp-carbon Raman fingerprint in the 1800-2300 cm$^{-1}$ range is related to the different collective stretching vibrations of sp-hybridized C–C, C≡C bonds [44]. Instead, the intense Raman band of PVA pellet closer to polyynes region is at about 2910 cm$^{-1}$, which refers to the symmetrical stretching vibrational mode $\upsilon_s(CH_2)$. It also shows two shoulders, one at 2851 cm$^{-1}$ related to weak intensity stretching mode $\upsilon(CH)$ and one at 2934 cm$^{-1}$ related to medium intensity asymmetrical stretching mode $\upsilon_a(CH_2)$ [45]. Such features, typical of the semi-crystalline polymer, can change due to structural disorder, as discussed below. SERS spectra of L-0.03%, L-0.5%, L-1%, L-3%, L-10% solutions and silver colloids (denoted by "L-x%_Ag") are presented in **Figure 2a**. These spectra are compared to the spectrum of the ablation in water (L-0%_Ag) and to the one of the liquid background ("L background"), composed by PVA1%. Two main bands in the polyynes region are visible in all the spectra: one at 2000-2200 cm$^{-1}$, due to collective CC stretching of the sp chain (effective conjugation coordinate (ECC) mode [46]), and the other one at 1800-2000 cm$^{-1}$. The observed displacement of this second CC stretching feature at lower wavenumbers can be associated to the interaction between sp chains and silver nanoparticles that can modify the electronic structure of the chains causing an equalization of the bond lengths, in agreement with our previous works [8, 33, 47]. The bands at 2000-2200 cm$^{-1}$ of all the samples, except for L-10%_Ag, are characterized by four narrow features, which refer to polyynes of different length. Raman peaks of longer polyynes are expected at lower wavenumber, in agreement with molecular modelling and experiments reported in previous works [46-48]. In SERS experiments, spectra are affected by the solvent and by the presence of metal nanoparticles hence further studies of size-selected polyynes in Ag/PVA matrix are needed to assign each band to polyynes of a specific length. Concerning the PVA band, we note that its intensity increases as its concentration in water, as shown in case of L-3%_Ag and L-10%_Ag.

In Ag/PVA/polyynes solutions up to the PVA concentration of 1 wt.% (i.e. L-0.03%_Ag, L-0.5%_Ag, L-1%_Ag), SERS features of sp-carbon are more intense with respect to samples synthesized in water (L-0%_Ag). Exceeding that concentration, as in case of L-3%_Ag and L-10%_Ag, SERS spectra show weaker and less resolved sp-carbon bands than in L-0%_Ag. These effects indicate a non-trivial role of PVA and can be possibly explained considering the liquid viscosity in the polyynes formation, as reported for PLAL in decalin [11]. Decalin has been reported to favour the synthesis of polyynes because the reactive species, i.e. $C_2$ radicals, hardly diffuse in the medium due to the high viscosity and are strongly confined in the plasma region. In this way, polyynes keep on growing up in length to $C_{30}$, before being terminated by hydrogen atoms, as reported by Inoue and co-workers [11, 12]. In their works, they used a mixture of two different conformations of decalin, the cis-decalin with a viscosity at 25 °C of 3.042 mPa·s and the trans-decalin with a value of 1.948 mPa·s, giving an average viscosity of 2.50 mPa·s, which is close to the one of PVA 1% (~ 2.53 mPa·s at 25 °C [49]). Consequently, viscous media, as decalin and PVA1%, can help the plasma confinement and hence the synthesis of the species. However, if the viscosity is too high, the overall yield of polyynes can be hindered by the solvent. In fact, the already formed end-capped chains, due to their low mobility in the viscous liquid, remain in the plasma region, and they may be decomposed and thermally degraded. Another effect that may decrease the polyyne formation yield in high viscous liquids is the slower release of the microbubbles created during the ablation which can partially shield the laser beam [50, 51]. In order to observe if we can balance the drop of polyynes signal in high viscous solvents (PVA >1%) by increasing the ablation time, we ablated graphite for longer time (30 minutes instead of 15 minutes) in a solution of PVA10% (~25.91 mPa·s at 25 °C [52]), which is the maximum concentration of PVA that can dissolve in water [53]. Increasing the ablation time, a slight increment of the intensity of polyyne SERS signal is recorded (see inset of **Figure 2)** but always showing one broad band without the narrow features observed in the other cases, as if the signal is not enough intense to show the different contributions. We underline that the intensity in the spectra of **Figure 2** is also strongly dependent on the SERS enhancement which results from the interaction between polyynes and metal nanoparticles and the formation of so-called hot spots. Hence, we cannot exclude that PVA may surround nanoparticles and polyynes preventing the direct adsorption of polyynes on Ag colloids. Several different effects, all related with the concentration of PVA, can indeed play a role in the non-trivial behaviour we observe.

SERS spectra of solid nanocomposites resulting from the solidification of the solutions with polyynes, silver colloid and PVA in water are shown in **Figure 2b**, herein indicated with S-x%_Ag (with x being the PVA concentration ranging from 1% to 10%). It was not possible to prepare a free-standing film starting from a PVA concentration below 0.5% (i.e. from sample L-0.03%_Ag) due to the limited quantity of PVA. All the spectra of the nanocomposites are characterized by two main broad bands in the polyyne region, as in the liquid case, displaying the fact that polyynes survive in the solidification process. S-1%_Ag sample shows the highest SERS enhancement as compared to the other free-standing films spectra, in agreement with what happens in solution. However, SERS features in the sp-carbon region are less resolved in the solid sample with respect to the liquid case.

To investigate the differences between the liquid and solid SERS analysis in **Figure 2c**, we focus on the SERS spectra of liquid and solid samples with PVA1% (i.e. L-1%_Ag and S-1%_Ag). Liquid sample (L-1%_Ag) displays four resolved narrow bands over 2000 $cm^{-1}$, fitted by 4 Lorentzian curves positioned at 2155, 2120, 2074, 2036 $cm^{-1}$. On the contrary, in the solid sample (S-1%_Ag) the sp-carbon features are broader bands without the well-defined sharp features of the liquid case. Such broad band can be fitted with 3 Lorentzians (centred at 2151, 2104, 2080 $cm^{-1}$) which are shifted compared to those of L-1%_Ag. This can be caused by the different SERS effect in liquid and solid [22, 38, 54]. In liquid, polyynes have mobility and can be adsorbed more easily on Ag nanoparticles leading to creation of hot spots which can be very sensitive to the specific polyyne lengths. Conversely, in a solid phase, polyynes and Ag colloids are fixed by the PVA matrix, and the different chemical environments of polyynes in proximity of metal nanoparticles may lead to a broad band, as observed in a previous work [33]. The band below 2000 $cm^{-1}$ appears as a broad band both in liquid and in solid samples and a detailed analysis of the SERS process is needed to properly analyse this band. By comparing liquid and solid sample with 1% PVA, we observe that in S-1%_Ag such band is more intense than the band at higher wavenumber, contrarily to what observed in L-1%_Ag. This effect can indicate that long polyynes or cumulenic complexes formed by the Ag interaction, which both give a contribution to the band at lower wavenumber [47], are more preserved by the solid polymeric matrix. The mechanism of SERS is still not completely understood, so further studies are needed to analyse the complex physical and chemical phenomena involved.

To better understand how the structure of the polymer changes with the presence of silver nanoparticles and polyynes, we compared the PVA CH stretching band in case of a pellet, a PVA film drop cast from water, aqueous solutions with Ag colloid and polyynes and the corresponding solid nanocomposites (see **Figure 2d**). We observe no differences between the PVA band in the pellet and in the film obtained from aqueous solution of PVA10%, both showing a main peak located at 2910 cm$^{-1}$, a shoulder at 2934 cm$^{-1}$ and a weak shoulder at 2851 cm$^{-1}$. Looking at L-10%_Ag, we notice that the main peak shifts up to 2925 cm$^{-1}$ but if the solution is dried, its position goes back to that of the pellet. Also in case of the aqueous solution of L-1%_Ag, the PVA peak moves to 2933 cm$^{-1}$, but in this case it shows a small shift to 2928 cm$^{-1}$ when the material is deposited; moreover, the shoulder at lower wavenumber become more visible and the other one at higher wavenumber is barely evident. It is known in literature that PVA band changes the position from solid pellet or cast film (2910, 2911 cm$^{-1}$), to aqueous solution (2918 cm$^{-1}$) [55]. The anomalously high peak wavenumber of the co-deposited L-1%_Ag film suggests that the presence of Ag colloid and polyynes could limit the capability of the polymer to form ordered crystalline phases during the solidification process. Interestingly, if we deposit from a solution where PVA concentration in water is low, e.g. PVA1%, we get a film S-1%_Ag, showing a CH stretching band position very close to that of PVA in water, namely to the wavenumber typical of conformationally disordered polymer chains. This indicates that polyynes, Ag nanoparticles and polymer form an intimate blend in the solid phase obtained. Instead, in case of higher concentration of PVA as in S-10%_Ag, the PVA band wavenumber matches the one of the pellet, which suggests that segregation of PVA crystalline domains occurs during the solidification. This effect can reduce the blending between all the components and encapsulation of polyynes.

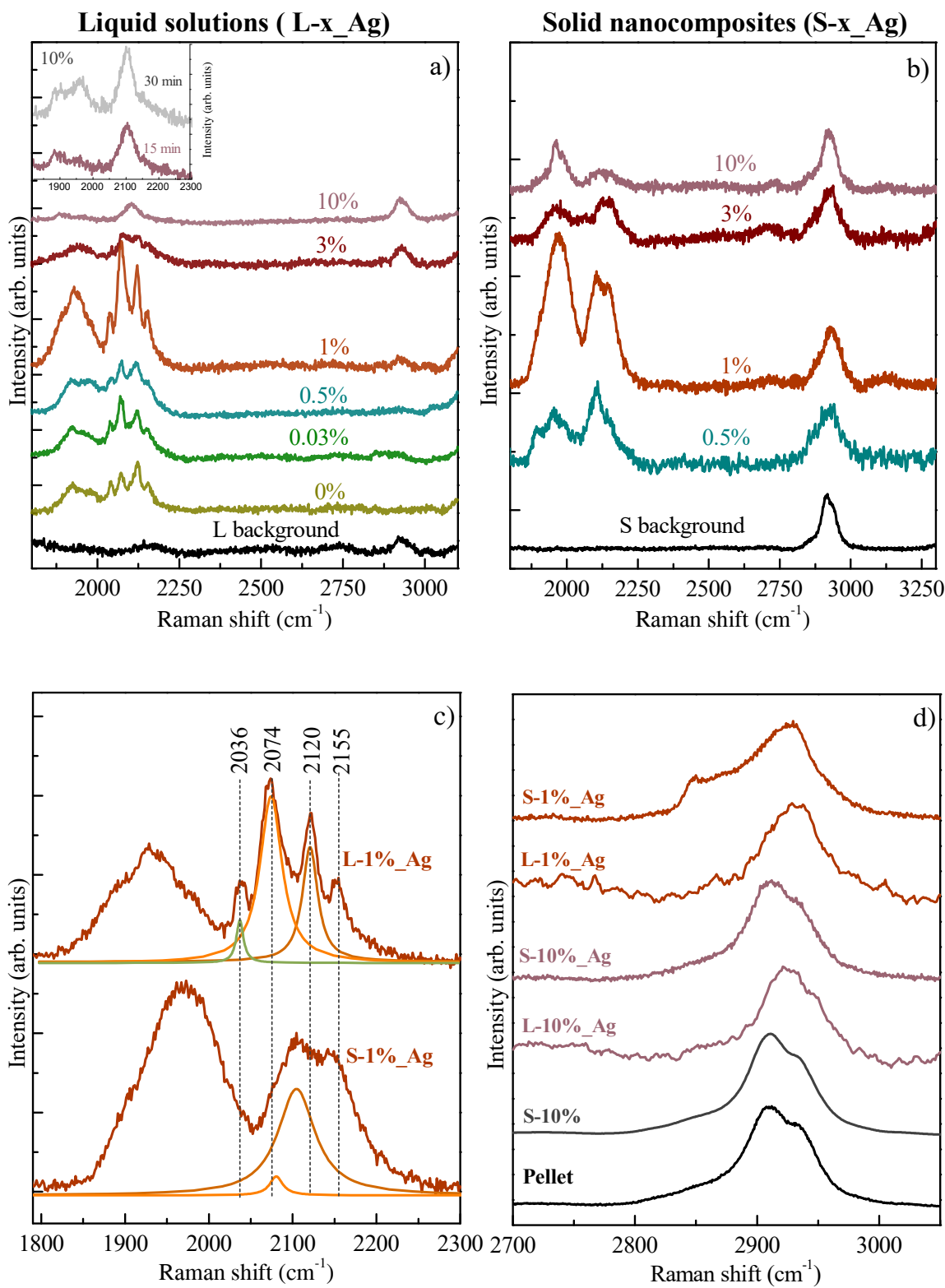

*Figure 2. a) SERS spectra of PVA/Ag/polyynes solutions at different concentration. In black line a spectrum of liquid PVA1% before ablation (L background). Inset: PVA10%/Ag/polyynes solutions spectra after 15 and 30 minutes of ablation. b) Raman spectrum of the PVA pellet in dotted black line and SERS spectra of PVA/Ag/polyynes free-standing films. In black, the free-standing film of solidified PVA1% without polyynes and Ag colloid (S background). c) SERS spectra of PVA1%/Ag/polyynes solution and the corresponding solid after Lorentz fitting. d) SERS spectra zoomed on the stretching vibrational mode of PVA in different phases.*

The free-standing films obtained after solvent evaporation of polyynes in PVA1% (**Figure 3a**) are characterized by a non-homogeneous distribution of silver nanoparticles which, during solvent evaporation, accumulate preferentially along the original drop edge due to the coffee ring effect. Such phenomenon can be visible by naked eyes, thanks to the colour difference from the centre to the edge of the sample. SEM images confirmed a higher concentration of Ag colloids at the film edge with respect to the centre, characterized by fewer nanoparticles but locally well dispersed (see **Figure 3b** and **c**). To investigate the effect of the inhomogeneity of the Ag nanoparticles in the sample, we mapped its SERS response along a line from the centre to the border, as drawn in **Figure 3a**. All the spectra have been normalized with respect to the PVA band. Polyynes signal is always present and its intensity decreases moving towards the edge of the film where Ag nanoparticles are more concentrated, as shown in **Figure 3d**. To evaluate this effect, we plot the ratio of the areas of the whole polyyne signal between 1800-2200 cm$^{-1}$ and the PVA peak as a function of the distance from the centre to the border of the sample (see inset of **Figure 3d**).

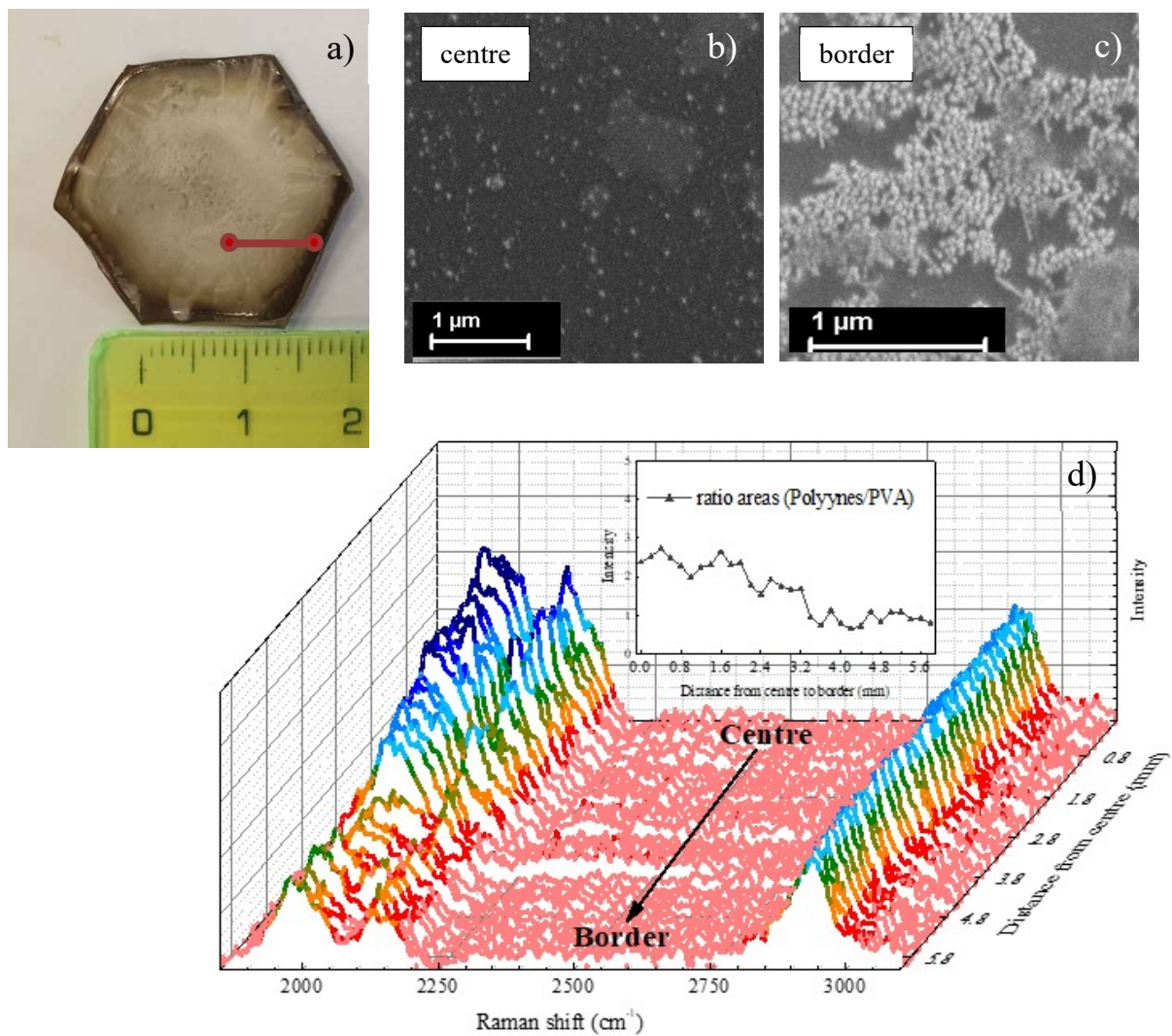

*Figure 3. a) Free-standing film of PVA/Ag/polyynes (S-1%_Ag), b) SEM image of the centre of the sample, c) SEM image of the border of the film, d) Normalized SERS spectra of the free-standing film from the centre to the border with step size of 200 µm following the line drawn on the Figure 3a). Inset: Ratio between area of the polyynes bands to the area of the PVA peak.*

The SERS signal of polyynes (1800-2250 cm$^{-1}$) lowers at the edge of the sample where the concentration of silver colloids is higher. If we assume a uniform distribution of polyynes, overabundant Ag nanoparticles may occupy the active site for SERS effect in place of sp-carbon chains leading to an overall lower polyyne signal at the edge than in the centre [56]. This result highlights how hard it is to obtain a good SERS enhancement because of its dependence on several factors including the concentration of Ag nanoparticles and the position of plasmonic peak which depends on the nanoparticle size and aggregation [33].

Once the films were prepared and the morphologically homogeneous areas with higher SERS enhancement were selected, we studied the stability of polyynes immobilized in the polymer matrix. We performed SERS analysis of the liquid and solid samples at different times starting from the as-prepared sample. **Figure 4** displays the comparison between the SERS time evolution of liquid solution of Ag colloid and PVA1% (L-1%_Ag) and the corresponding solid nanocomposite (S-1%_Ag). Spectra are normalized with respect to the peak of PVA at 2910 cm$^{-1}$, which does not overlap to Raman features of any other species. In the case of the liquid, the well-defined polyyne features of the SERS spectrum are degraded in two broad bands after already 1 week and, after 2 months, no features are visible in the sp-carbon spectral region. After 3 weeks, the reduction of polyynes in PVA/Ag solution, evaluated considering the integrated area of the SERS signal in sp-carbon range, is of about 60%, which is anyway strongly reduced with respect to the case of polyynes in aqueous Ag colloid, whose area reduction after 3 weeks is of 95% (as reported in [18]). In the spectra of solid nanocomposite, we observe after 1 week a decrease of the intensity of the polyynic SERS bands of the order of 50% but then, the spectrum does not substantially change up to 11 months. Our results agree with the few works addressing the stability of polyynes in PVA. Okada et al. showed that Ag-polyynes aggregates in PVA are stable for a month [37]. An and co-workers observed SERS signal from Au/polyynes/PVA films up to 6 months [38]. In our work, we indicate that stability of polyynes in liquid media is improved in presence of PVA, even if only slightly, because the reactions between sp-carbon chains are partially inhibited by the presence of the polymer. However, sp-carbon chains are not completely immobilized in PVA solution and the mechanism of crosslinking does not stop completely. In case of solid nanocomposite, the stabilizing effect is instead quite clear and effective. In fact, once polyynes are encapsulated in PVA matrix, they are somehow protected, and they are stable for a prolonged time.

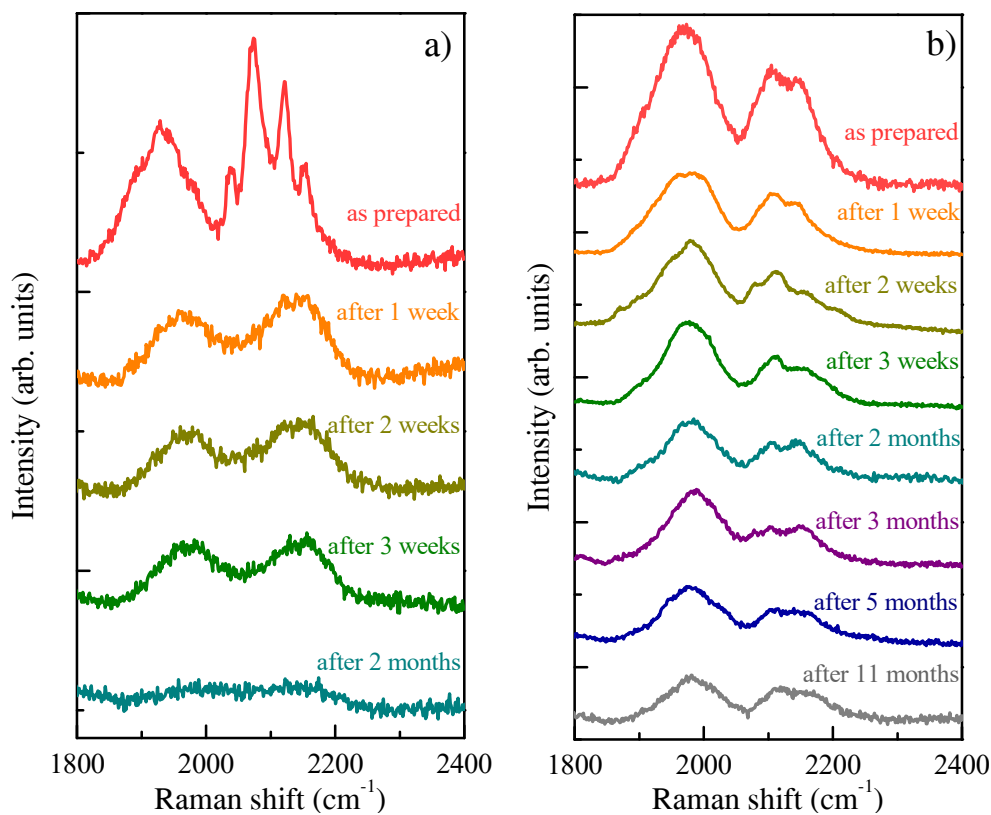

**Figure 4**. Evolution in time of SERS spectra: **a)** solution of PVA1% with Ag colloid (L-1%_Ag) and **b)** free-standing nanocomposite of PVA1%/Ag/polyynes (S-1%_Ag).

## 3. Conclusions

In this work, we have discussed the ablation of graphite directly in a pure aqueous solution of PVA, as an in situ method to form polyynes in a polymer matrix. In this way, we avoided diluting polyynes by the subsequent addition of the polymer in the pristine solutions or to degrade them by the heating step necessary when PVA granules are directly dissolved in the solution containing sp-carbon chains. This study has involved the preparation of different PVA/polyynes solutions varying the PVA concentration in water. After the addition of Ag colloid to aqueous solution of PVA/polyynes and its drying, we obtained a polyynes-based free-standing film. We have observed that aqueous PVA, as a solvent in PLAL, can give a two-fold advantage with respect to the ablation in pure water. From one side, it favours polyynes formation as noticeable from the increased UV-Vis polyynes absorption. From the other, we found increasingly resolved SERS spectra with the growth of PVA concentration up to 1 wt.%, both for Ag/PVA/polyynes solutions and the corresponding nanocomposites. Moreover, the combined SEM and SERS map analysis on free-standing films allowed us to individuate the area of the nanocomposite with homogeneous morphology and the highest SERS intensity, where it was possible to perform a stability study. Periodic SERS measurements proved that hydrogen-capped polyynes embedded in Ag nanoparticles and PVA are still stable after 11 months. In this way, our simple method of synthesis of hydrogen-capped polyynes in PVA and the enhanced stability in the polymer matrix is promising for future optical and mechanical characterization of polyyne-based film.

# 5. Acknowledgements

Authors acknowledge funding from the European Research Council (ERC) under the European Union's Horizon 2020 research and innovation program ERC-Consolidator Grant (ERC CoG 2016 EspLORE grant agreement No. 724610, website: www.esplore.polimi.it).